\documentclass[12pt]{iopart}

\usepackage{iopams}
\usepackage[dvips]{graphicx}
\usepackage{epsf}
\usepackage{rotating}

\newcommand{\bd     }{\begin{displaymath}}
\newcommand{\ed     }{\end{displaymath}}

\newcommand{\s      }{\sigma}

\newcommand{\be     }{\beta}

\newcommand{\bra    }{\langle}
\newcommand{\ket    }{\rangle}
\newcommand{\Bra    }{\left\langle}
\newcommand{\Ket    }{\right\rangle}

\newcommand{\la     }{\lambda}

\newcommand{\bA     }{\mbox{\boldmath$A$}}

\newcommand{\bn     }{\mbox{\boldmath$n$}}

\renewcommand{\bs     }{\mbox{\boldmath$s$}}

\newcommand{\bz     }{\mbox{\boldmath$z$}}

\newcommand{\bx     }{\mbox{\boldmath$x$}}
\newcommand{\by     }{\mbox{\boldmath$y$}}
\renewcommand{\br     }{\mbox{\boldmath$r$}}
\newcommand{\bt     }{\mbox{\boldmath$t$}}

\newcommand{\mcA    }{\mathcal{A}}
\newcommand{\mcI    }{\mathcal{I}}

\begin{document}
\title{Average and reliability error exponents in low-density parity-check
           codes}
\author{ N.S.~Skantzos$^{\dagger *}$, J.~van Mourik$^{\dagger}$,
             D.~Saad$^{\dagger}$  and Y.~Kabashima$^{\ddagger}$}
\address{ $^\dagger$Neural Computing Research Group,
              Aston University, Birmingham B4 7ET, UK }
\address{ $^{*}$ Institut for Theoretical Physics, Celestijnenlaan 200D,
KULeuven, Leuven, B-3001 Belgium}
\address{$^\ddagger$Dept.\@ of Computational Intelligence \&
              Systems Science, Tokyo Institute of Technology,
              Yokohama 2268502, Japan }


\begin{abstract}
We present a theoretical method for a {\em direct} evaluation of the
average and reliability error exponents in low-density parity-check 
error-correcting
codes using methods of statistical physics.  Results for the binary
symmetric channel (BSC) are presented for codes of both finite and
infinite connectivity.
\end{abstract}


\section{Introduction}
\label{sec:intro}

Low-density parity-check codes (LDPC) have attracted significant
interest in recent years due to their simplicity and exceptionally
high performance~\cite{Richardson}. Their simplicity and inherent
randomness make them amenable to analysis using established
methods in the area of statistical physics. These have been
employed in a number of papers~\cite{renato}-\cite{franz} to gain
insight into the properties of LDPC codes and to evaluate their
performance.

These studies include the evaluation of critical noise levels for
given codes~\cite{renato}, an exact calculation of weight and
magnetisation enumerators~\cite{jort}, the performance of irregular
codes~\cite{irregular}, properties of codes in real-valued
channels~\cite{tanaka}, and the derivation of bounds for the
reliability exponent~\cite{yoshiyuki}, to name but a few. These
studies also represent the interdisciplinary nature of this research
area and illustrate the successful interaction between researchers in
the two disciplines.

The evaluation of error exponents has been a long-standing problem in
information theory~\cite{gallager_book,viterbi}. Efforts to obtain
exact expressions and/or bounds to the error exponent resulted in
partial success; although tight bounds have been derived in the case
of random codes and LDPC with infinite
connectivity~\cite{gallager_book}, only limited results have been
obtained for sparely connected codes.  Main stream techniques to
tackle the problem include sphere-packing and union-bound
arguments~\cite{viterbi,gallager_book}.  Below a certain code-rate
value, the estimated bounds also become loose and require using the
`expurgated exponent' techniques~\cite{gallager_book} for obtaining a
tighter bound.

In this paper, we employ methods of statistical physics to
evaluate directly the average error exponent and typical
reliability exponent in Gallager and MN~\cite{MN} LDPC codes. The
{\em average error exponent} is obtained by carrying out averages
over the ensemble of randomly generated LDPC codes of given rate
and connectivity; while the {\em reliability exponent} is obtained
by selecting the best codes in that ensemble. Averages result in
the emergence of macroscopic variables, representative of the
ensemble properties, that can be obtained numerically and used to
calculate the average error exponent (in the current calculation
we assume that short loops, which contribute polynomially to the
block error probability in LDPC codes~\cite{miller}, have been
removed). Average error exponent solutions have been obtained for
both finite and infinite connectivity vector ensembles, while
reliability exponent solutions have been obtained only in the case
of infinite connectivity.

As a reference point to test our theory, we use known results obtained
in the information theory literature for solvable limits (e.g. codes
of infinite connectivity), and find that our method reproduces them
exactly. Perhaps not surprisingly, we also find that at fixed noise
level and code rate, the reliability exponent for codes of finite
connectivity is always upper-bounded by that of the infinite-connectivity
case.

Before we proceed, the distinction between the typical bounds found
previously using methods of statistical physics~\cite{yoshiyuki}, and
the current calculation should be clarified. In the former, one
employs methods of statistical physics to calculate the typical value
of a {\em bound} based on inequalities introduced by Gallager; while
in the current calculation, a direct estimation of the average error
exponent, rather than a bound, is sought. An additional advantage of
the current approach is that it can be extended to provide
reliability exponent values for LDPC codes by restricted averages
over codes of high performance.

The paper is organised as follows: In section~\ref{sec:definitions},
we introduce the general coding framework and the technique used.  In
sections~\ref{sec:solution} and~\ref{sec:exactsolution} we present an
outline of the derivation and the solutions obtained in both finite
and infinite connectivity cases respectively. In Section~\ref{sec:MN}
we compare the error exponent results obtained for MN codes to
those of Gallager codes in both finite and infinite connectivity
cases.  Discussion and conclusions are presented in
section~\ref{sec:conclusions}.

\section{Definitions}
\label{sec:definitions}

A regular ($k,j$) Gallager error-correcting code is defined by the binary
$(N-K)\times N$ (parity check) matrix $A=[C_1|C_2]$, which is known to both
sender and receiver. The $(N-K)\times(N-K)$ matrix $C_2$ is taken to be
invertible. The number of non-zero elements in each row of $A$ is given by
$k$, while the number of non-zero elements per column is given by
$j\equiv k(N-K)/N$.

Gallager's encoding scheme consists of generating a codeword
$\bt\in\{0,1\}^N$ from an information (message) vector
$\bs\in\{0,1\}^K$ (with $N>K$) via the linear operation $\bt=G^T\bs$ (mod 2)
where $G$ is the generator matrix defined by $G=[I| C_2^{-1}C_1]$ (mod 2).
The code rate is then given by $R\equiv K/N=1-j/k$, and measures the
information redundancy of the transmitted vector.

Upon transmission of the codeword $\bt$ via a noisy channel, taken here
to be a BSC, the vector $\br=\bt+\bn^0$ (mod 2) is received, where
$\bn^0\in\{0,1\}^N$ is the true channel noise.  The statistics of the BSC
is fully determined by the flip rate $p\in[0,1]$:
\begin{equation}
P(n^0_i)=(1-p)\,\delta_{n^0_i,0}+p\,\delta_{n^0_i,1}
\label{noise}
\end{equation}
Decoding is carried out by multiplying $\br$ by $A$ to produce the syndrome
vector $\bz=A\br=A\bn^0$, since $AG^{T}=0$ by construction. In order to
reconstruct the original message $\bs$, one has to obtain an estimate $\bn$
for the true noise $\bn^0$. First we select the parity check set of $A$ and
$\bn^0$, i.e. all $\bn$ that satisfy the parity check equations:
$\mcI_{pc}(A,\bn^0)\equiv\{\bn~|~A\bn=A\bn^0\}$. Since all operations are
performed in modulo 2 arithmetic, $\mcI_{pc}(A,\bn^0)$ typically contains
$\exp[NR\ln(2)]$ candidates for the true noise vector $\bn^0$.

It was shown (see e.g. \cite{renato,yoshiyuki,montanari} for technical
details) that this problem can be cast into a statistical mechanics
formulation, by replacing the field $(\{0,1\},+{\rm mod(2)})$ by
($\{1,-1\},\times$), and by adapting the parity checks
correspondingly. From the parity check matrix $A$ we construct the
binary tensor $\mcA=\{\mcA_{\bra{i_1}\cdots{i_k}\ket},
1\!\le\!i_1\!<\!i_2\cdots<\!i_k\!\le\!N\}$, where
$\mcA_{\bra{i_1}\cdots{i_k}\ket}=1$ if $A$ has a row in which the
elements $\{i_c, c=1,\ldots, k\}$ are all 1 (i.e. when the bits $\bra
i_1\cdots{i_k}\ket$ are involved in the same parity check), and 0
otherwise. The fact that each bit $i_1=1,\ldots, N$ is involved in
exactly $j$ parity checks is then expressed by $\sum_{i_2<\cdots<i_k}
\mcA_{\bra i_1\cdots i_k\ket}=j,~~\forall\ i_1=1,\ldots,N$
and the parity check equations become $\prod_{c=1}^k n_{i_c}=\prod_{c=1}^k
n^0_{i_c}$,  $~~\forall \mcA_{\bra i_1\cdots i_k\ket}=1$.

Decoding now consists in selecting an $\bn$ from $\mcI_{pc}(\mcA,\bn^0)$, on the
basis of its noise statistics, which are fully described by its {\em
magnetisation} $m(\bn)={1/N}\sum_i n_i$ (corresponding to the weight in the
information theory literature). Note that the number of flipped bits in a
candidate noise vector $\bn$ is given by $N(1-m(\bn))/2$. Therefore, we
introduce a Hamiltonian or cost function for each noise candidate that is
negatively proportional to its magnetisation:
\begin{equation}
H(\bn)=-F\sum_{i=1}^{N} n_i=-FNm(\bn)
\end{equation}
where we take $F=\frac12\log\frac{1-p}{p}$, such that up to
normalisation $\exp(-H(\bn))$ yields the correct prior for candidate
noise vectors generated by the BSC \cite{nishimori_book}. Then, a
vector $\bn$ from $\mcI_{pc}(\mcA,\bn^0)$ with the highest
magnetisation (lowest weight) is selected as a solution; this
corresponds to Maximum A Posteriori (MAP) decoding.

We are now interested in the probability that other candidate noise
vectors are selected from the parity check set
$\mcI_{pc}(\mcA,\bn^0)$, other than the correct (i.e. true) noise
vector $\bn^0$, for any given combination $\{\bn^0,\mcA\}$; this is
termed the {\em block error probability}. In order to calculate this
probability, we introduce an indicator function:
\begin{equation}
\Delta(\bn^0,\mcA)=\lim_{\beta_{1,2}\to\infty}\lim_{\lambda_{1,2}\to\pm\la}
          \left.\left[Z_1^{\lambda_1}(\bn^0,\mcA;\beta_1)~
                      Z_2^{\lambda_2}(\bn^0,\mcA;\beta_2)
          \right]\right|_{\beta_1=\beta_2=\beta}
\label{eq:Delta_def}
\end{equation}
where
\begin{equation}
Z_1(\bn^0,\mcA;\beta_1)=\hspace*{-6mm}\sum_{\bn\in\mcI_{pc}(\bn^0,\mcA)
   \backslash\bn^0}
   \hspace*{-6mm}e^{-\beta_1 H(\bn)},
\hspace{10mm}
Z_2(\bn^0,\mcA;\beta_2)=\hspace*{-6mm}\sum_{\bn\in \mcI_{pc}(\bn^0,\mcA)}
   \hspace*{-6mm}e^{-\beta_2 H(\bn)}.
\label{eq:partition_sums}
\end{equation}
The partition functions $Z_1(\bn^0,\bA;\beta_1)$ and
$Z_2(\bn^0,\bA;\beta_2)$ differ only in the exclusion of $\bn^0$ from
$Z_1$. If the true noise $\bn^0$ has the highest magnetisation of all
candidates in the parity check set (decoding success), the Boltzmann
factor $\exp[-\beta H(\bn^0)]$ will dominate the sum over states in
$Z_2$ in the limit $\beta\to\infty$, and $\Delta(\bn^0,\mcA)=0$.
Alternatively, if some other vector $\bn\neq\bn^0$ has the highest
magnetisation of all candidates in the parity check set (decoding
failure), its Boltzmann factor will dominate both $Z_1$ and $Z_2$ and
$\Delta(\bn^0,\mcA)=1$. Note that the separate temperatures $\beta_1$
and $\beta_2$, which are put to be equal to $\be$ in the end, and the
powers $\lambda_{1,2}$ which are taken to be $\pm\lambda$ in the end,
have been introduced in order to allow us to determine whether
obtained solutions are physical or not. The power $\lambda \ge 0 $
have been introduced to restrict the indicator function results to
0/1. In principle, this can be done by taking the limit $\lambda
\rightarrow 0$; however, in section~\ref{sec:solution}, we show that
finite $ 0<\lambda <1 $ values will be used due to various
constraints.

To derive the {\em average error exponent}, we take the logarithm of the
indicator function average with respect to all possible realisations of true
noise vectors $\bn^0$, and the ensemble of regular $(k,j)$ codes $\mcA$:
\begin{equation}
Q=\lim_{N\to\infty}\frac{1}{N}\log
  \left\bra\left\bra\Delta(\bn^0,\mcA)\right\ket_{\bn^0}\right\ket_{\mcA}
\label{eq:reliability}
\end{equation}
where
\begin{equation}
\bra f(\bn^0)\ket_{\bn^0}=\frac{1}{(2\cosh{F})^N}\sum_{\bn^0}
 \exp(F\sum_in_i^0)~f(\bn^0)
\label{eq:truenoise_average}
\end{equation}
and
\begin{equation}
\bra f(\mcA)\ket_\mcA=
  \frac{\sum_{\mcA}\prod_{i_1=1}^N
     \delta[\sum_{{i_2}<\cdots<{i_k}}\mcA_{\bra i_1\cdots i_k\ket}-j]~f(\mcA)}
       {\sum_{\mcA}\prod_{i_1=1}^N
     \delta[\sum_{{i_2}<\cdots<{i_k}}\mcA_{\bra i_1\cdots i_k\ket}-j]        }.
\label{eq:tensor_average}
\end{equation}

To obtain an expression for the {\em reliability exponent} one carries out a
similar calculation with one main difference: prior to averaging the  indicator
function  over the  ensemble of regular $(k,j)$ codes $\mcA$, one takes the
averaged expression with respect to realisations of true noise vectors $\bn^0$
to a power $r$ which favours code constructions with a low average error
probability (i.e., $r<1$). The logarithm of the expression averaged over the
ensemble of codes $\mcA$ is then divided by $r$ to remove the exponent. The
expression calculated is:
\begin{equation}
Q_{r}=\lim_{N\to\infty}\frac{1}{Nr}\log
  \left\bra \left[ \left\bra\Delta(\bn^0,\mcA)\right\ket_{\bn^0}
  \right]^{r}\right\ket_{\mcA}
\label{eq:reliability_exp}
\end{equation}
Since there are only discrete degrees of freedom, physically meaningful
solutions must have a non-negative entropy, requiring the disorder-averaged
entropies of the two partition functions (\ref{eq:partition_sums}) to be
non-negative. 
Note that due to the order of taking the logarithm vs the various averages, 
expressions (\ref{eq:reliability}) and (\ref{eq:reliability_exp}) 
are not equivalent to a (quenched) disorder-averaged free energy. 
Using general principles one can show that 
for general values of $\beta_{1,2}$ and $\lambda_{1,2}$, 
the disordered-averaged entropies (with averages taken over 
the joint distribution of code-constructions $\{\mcA\}$, true- and 
candidate-noise $\{\bn^0,\bn\}$ as suggested by 
(\ref{eq:reliability}) and (\ref{eq:reliability_exp})) are given, 
for both calculations
(\ref{eq:reliability}) and (\ref{eq:reliability_exp}), by
\begin{equation}
\bra S_x\ket=\frac{\partial Q_r}{\partial \lambda_x}-
                 \frac{\beta_x}{\lambda_x}
                 \frac{\partial Q_r}{\partial \beta_x}\geq 0,\hspace*{1cm}x=1,2
\label{eq:entropies}
\end{equation}
which have to be positive.

\section{Average error exponent - general solution}
\label{sec:solution}

Using standard statistical physics methods such as in \cite{nishimori_book},
we perform the gauge transformation $n_i\to n_i n_i^0$, and the averages over
true noise (\ref{eq:truenoise_average}) and code constructions
(\ref{eq:tensor_average}). In the case of $r\neq 1$, each quantity
carries two indices (a replica index and another index coming from the
power $r$); however, the two indices factorise {\em unless an
explicit, more complex, symmetry breaking structure is
introduced}. Here, we do not assume a more complex structure that
entangles the two types of indices; we also assume the simplest
replica symmetric scheme \cite{wong-sherrington} to arrive at the
following expression for the average error exponent ($r=1$), and for
the reliability exponent (optimised $r$):
\begin{equation}
Q_{r}(\beta_1,\beta_2,\lambda_1,\lambda_2)=\frac{1}{r}
   {\rm Extr}_{\pi,\hat{\pi}}\left[
   \frac{j}{k}\log~I_1[\pi]-j\log~I_2[\pi,\hat{\pi}]+\log~I_3[\hat{\pi}]
                             \right]
\label{eq:reliability_final_r}
\end{equation}
where
\begin{equation}
I_1=\int\prod_{c=1}^k\left\{d\pi(x_c,y_c)\right\}
\left(\frac{1+\prod_{c=1}^k x_c}{2}\right)^{r\lambda_1}
\left(\frac{1+\prod_{c=1}^k y_c}{2}\right)^{r\lambda_2}
\label{eq:I1_r}
\end{equation}
\begin{equation}
I_2=\int\left\{ d\pi(x,y)\ d\hat{\pi}(\hat{x},\hat{y})\right\}
  \left(\frac{1+x\hat{x}}{2}\right)^{r\lambda_1}
  \left(\frac{1+y\hat{y}}{2}\right)^{r\lambda_2}
\label{eq:I2_r}
\end{equation}
\begin{eqnarray}
I_3
 &=&\int\prod_{c=1}^j\left\{d\hat{\pi}(\hat{x}_c,\hat{y}_c)\right\}
  \left\bra
  \left[\sum_{u=\pm 1}e^{\beta_1 Fn^0 u}\prod_{c=1}^j
    \left(\frac{1+u\hat{x}_c}{2}\right)\right]^{\lambda_1}\right.
\nonumber \\ &&\hspace*{37mm}\times\left.
    \left[\sum_{v=\pm 1}e^{\beta_2 F n^0 v}\prod_{c=1}^j
    \left(\frac{1+v\hat{y}_c}{2}\right)\right]^{\lambda_2}
  \right\ket_{n^0}^r
\label{eq:I3_r}
\end{eqnarray}
where we have used the short-hand notation $df(x,y)$=$dxdy~f(x,y)$.
For $r=1$, functional extremisation of (\ref{eq:reliability_final_r}) with
respect to the densities $\pi(x,y)$ and $\hat{\pi}(\hat{x},\hat{y})$ results in
a closed set  of equations (reminiscent of `density evolution' equations
\cite{Richardson}):
\begin{equation}
\hat{\pi}(\hat{x},\hat{y})=
  \int\prod_{c=1}^{k-1}\left\{d\pi(x_c,y_c)\right\}
  \delta\left[\hat{x}-\prod_{c=1}^{k-1}x_c\right]\
  \delta\left[\hat{y}-\prod_{c=1}^{k-1}y_c\right]
\label{eq:SP1}
\end{equation}
\begin{equation}
\pi(x,y)=\frac{\Bra\!\Bra
     \delta\left[x-\frac{D_-(\hat{\bx};\beta_1)}{D_+(\hat{\bx};\beta_1)}\right]
     \delta\left[y-\frac{D_-(\hat{\by};\beta_2)}{D_+(\hat{\by};\beta_2)}\right]
               \Ket\!\Ket'}
              {\Bra\!\Bra1\frac{}{}\Ket\!\Ket'}.
\label{eq:SP2}
\end{equation}
where
\begin{eqnarray}
\Bra\!\Bra\cdot\frac{}{}\Ket\!\Ket'&\equiv&
\int\prod_{c=1}^{j-1}\left\{d\hat{\pi}(\hat{x}_c,\hat{y}_c)\right\}
\Bra D_+^{\lambda_1}(\hat{\bx};\beta_1)D_+^{\lambda_2}(\hat{\by};\beta_2)~\cdot~
\Ket_{n^0}, \\
D_\pm(\bz;\beta)&\equiv&[e^{\beta F n^0}\prod_{c=1}^{j-1}(1+z_c) ]
       \pm[e^{-\beta F n^0}\prod_{c=1}^{j-1}(1-z_c)].
\end{eqnarray}
For given $(\beta_1,\beta_2,\lambda_1,\lambda_2)$ in general, solutions to
(\ref{eq:SP1}) and (\ref{eq:SP2}) can only be obtained numerically.
Inserting these solutions into (\ref{eq:reliability_final_r}) we then
obtain $Q(\beta_1,\beta_2,\lambda_1,\lambda_2)$, which becomes the {\em average
 error exponent} for $\lambda_1=-\lambda_2=\lambda>0$, and for
$\beta_1=\beta_2=\beta\to\infty$.

We must recall, however, that physically meaningful solutions
must satisfy the conditions (\ref{eq:entropies}) stating that the entropies
related to the full and the restricted partition sums are non-negative.

We restrict ourselves to regions below the thermodynamic transition
where the average case is dominated by the ferromagnetic solution,
such that we can fix 
the system described by $Z_2$ in (\ref{eq:partition_sums})
to the ferromagnetic
solution.  This dominance is guaranteed if the following constraint is
satisfied
\begin{equation}
\left.\frac{\partial Q}{\partial \beta }
\right|_{\lambda_1=-\lambda_2=\lambda}\leq 0 \ .
\label{eq:condition3}
\end{equation}

It turns out that for given $\lambda>0$, the largest value of $\beta$ for which
(\ref{eq:condition3}) is satisfied is given by the simple expression
$\beta=1/(1+\lambda)$. Hence, in order to maximise $\beta$, we must look for the
smallest value $\lambda_*$ that satisfies the conditions on the non-negativity
of the entropies (\ref{eq:entropies}). Unfortunately, in general this value
$\lambda_*$ can only be obtained numerically. The value obtained for the average
error exponent by this analysis is then given by $Q(1/(1+\lambda_*),1/
(1+\lambda_*),\lambda_*,-\lambda_*)$ from (\ref{eq:reliability_final_r}).

\begin{figure}[t]
\vspace{-15mm}
\setlength{\unitlength}{1.4mm}
\begin{picture}(120,55)
\put( 10,  0){\epsfysize=40\unitlength\epsfbox{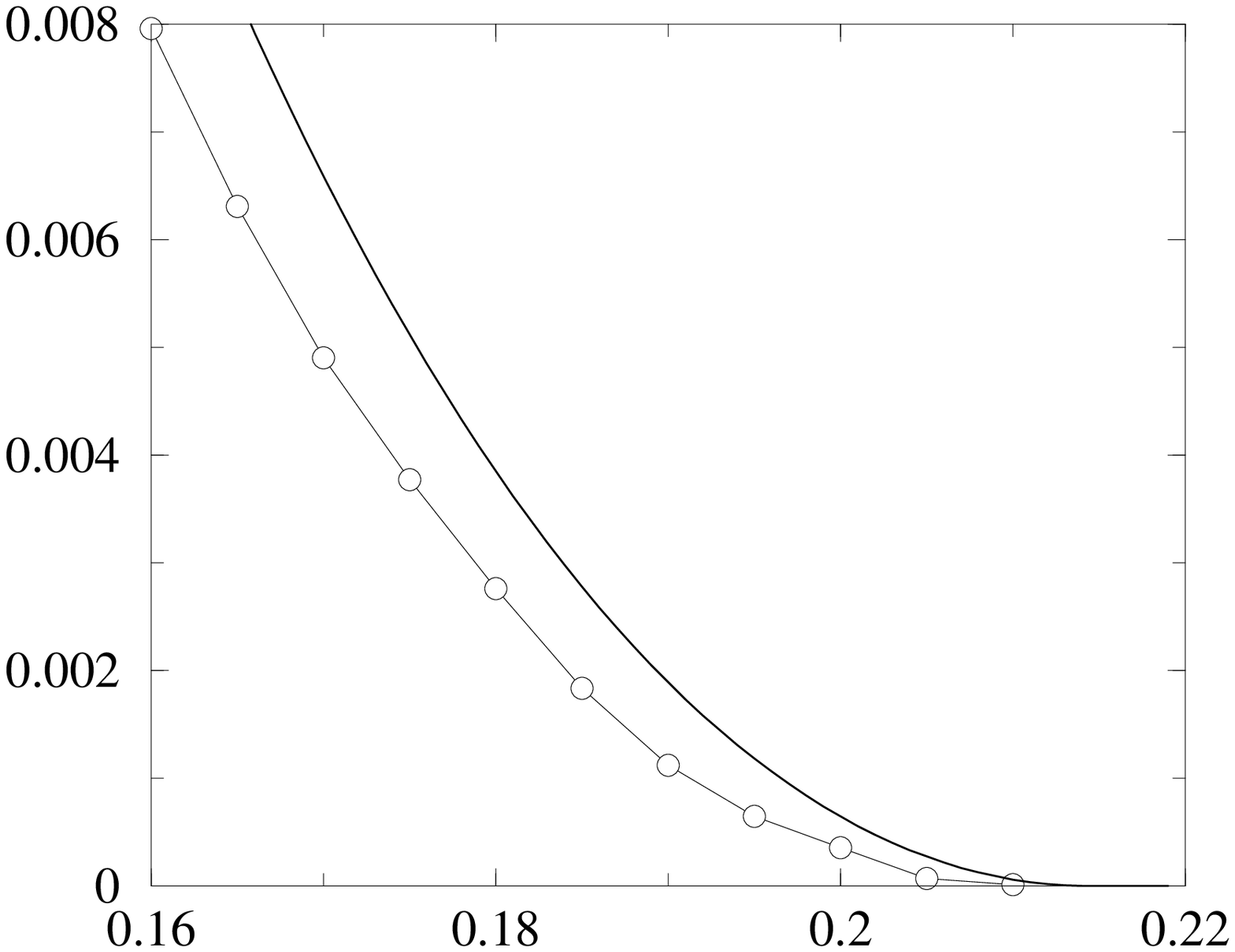}}
\put( 65,  0){\epsfysize=40\unitlength\epsfbox{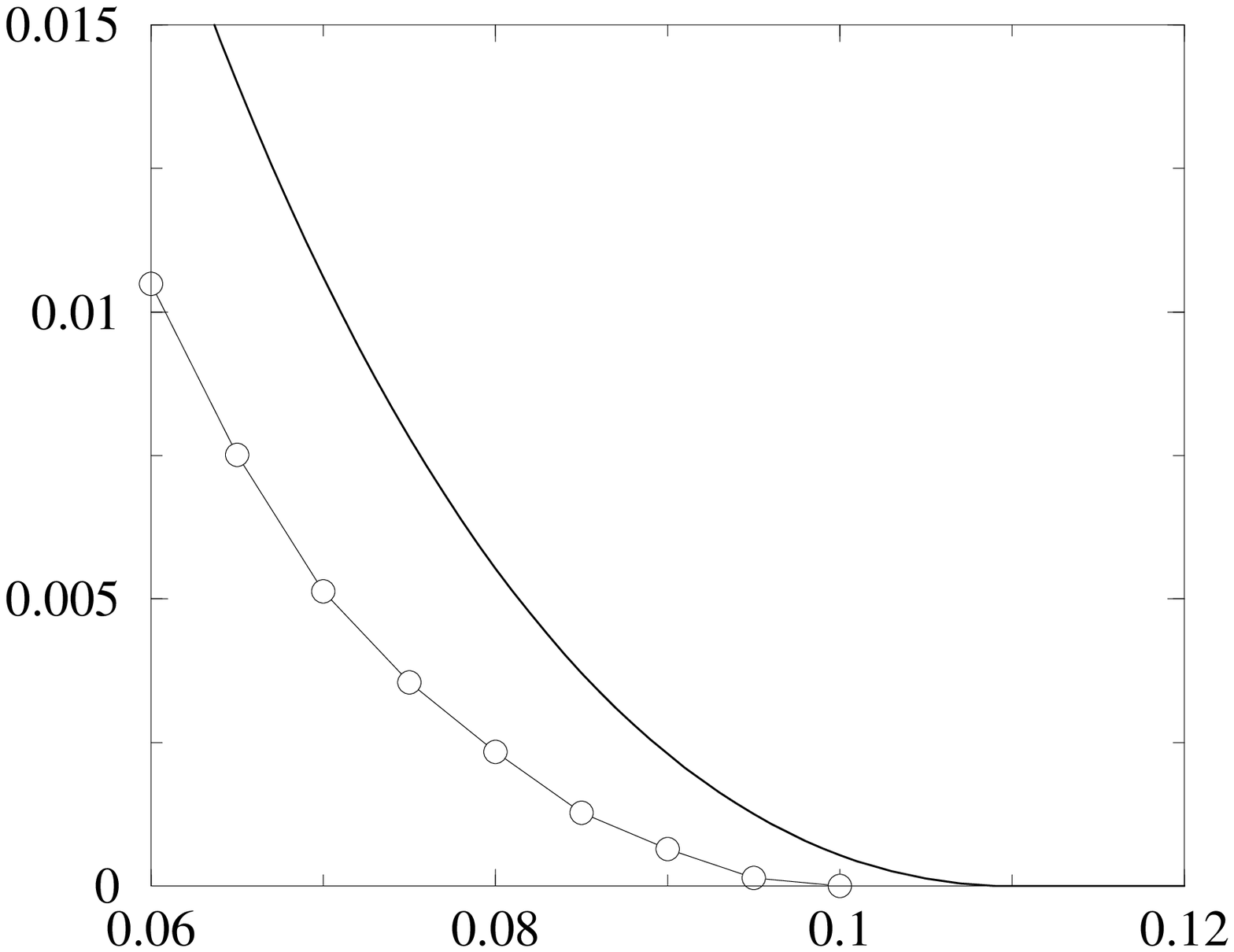}}
\put(5,25){$Q$}
\put(62,25){$Q$}
\put(40,0){$p$}
\put(95,0){$p$}
\end{picture}
\caption{ Average error exponent $Q$ as function of the flip rate $p$ for codes
 of $(k,j)=(4,3)$ (left picture) and $(k,j)=(6,3)$ (right picture). 
 Lines with markers
 correspond to the finite $(k,j)$ cases. 
 For comparison we also present (thick solid lines) the value of the
 average error exponent in the case of $k,j\to\infty$ with $R=1/4$ (left) and
 $R=1/2$ (right) as described in the analysis of section \ref{sec:exactsolution}.
 Note that the transition from type I to type II solution
 occurs at small $p$ values outside the range of this figure.
}
\label{fig:KCfinite}
\end{figure}

In figure \ref{fig:KCfinite} we present the obtained average error
exponent as a function of the flip rate for $(k,j)=(4,3)$, ($R=1/3$)
and $(k,j)=(6,3)$ ($R=1/2$) codes. We observe that the error exponent
indeed converges to zero, as it should, when the flip rate approaches
its critical value.

Notice the similarity between the equations obtained here and
in~\cite{yoshiyuki} in spite of the different starting points. It has
been shown in~\cite{yoshiyuki} that the analysis should be refined in
low rate regions by considering a more complex bound. The refined
analysis resulted in tight bounds of the error exponent even in the
region of low code-rates, similar to those obtained using expurgated
exponent methods. In the next section we will show that the selection
of 'best codes' through the optimisation of the power $r$, in
calculating the reliability exponent, provides similar results to
those obtained in~\cite{yoshiyuki}.

\section{An exactly solvable limit: $k,j\to\infty$}
\label{sec:exactsolution}

Whereas for finite density codes solutions for the average error exponent are
obtained numerically, in the limit of $k,j\to\infty$ (while keeping the rate
$R=1- j/k$ finite) one obtains two types of analytic solutions to equations
(\ref{eq:SP1}) and (\ref{eq:SP2}), which can be verified by
substitution. Moreover, in this limit one also obtains solutions in the
reliability exponent calculation (\ref{eq:reliability_exp}), which are generally
difficult to obtain for finite $k$ and $j$ values.

\subsection{Average error exponent}
Solutions obtained in the average error exponent calculation take the
following form:
\newline
Type I:
\begin{eqnarray}
\pi(x,y)&=&                  \frac12
\left[\delta(x-1)+\delta(x+1)\right]~\delta(y-1)
\nonumber\\
\hat{\pi}(\hat{x},\hat{y})&=&\frac12
\left[\delta(\hat{x}-1)+\delta(\hat{x}+1)\right]~\delta(\hat{y}-1)
\label{eq:typeI}
\end{eqnarray}
Type II:
\begin{eqnarray}
\pi(x,y)&=&\delta(y -1)~\Biggl[
         G_+(F(1\!+\!\beta_2\lambda_1))~\delta(x\!-\!\tanh(\beta_1 F)) \nonumber\\
        &+&  G_-(F(1\!+\!\beta_2\lambda_2))~\delta(x\!+\!\tanh(\beta_1 F))
 \Biggr]
\nonumber\\
\hat{\pi}(\hat{x},\hat{y})&=&\delta(\hat{y}-1)~\delta(\hat{x})
\label{eq:typeII}
\end{eqnarray}
with $G_\pm(x)=\frac12[1\pm\tanh(x)]$.

Taking $\beta_1=\beta_2=\beta$ and $\lambda_1=-\lambda_2=\lambda$, the average
error exponent as obtained from the type I solution is given by
\begin{equation}
Q_{I}=-\frac{j}{k}\log 2 -\log \cosh F+\log \cosh(\beta F\lambda)
       +\log 2 \cosh(F-\beta F\lambda) \ .
\end{equation}
We find that the entropies (\ref{eq:entropies}) are always identically zero,
and that the constraint (\ref{eq:condition3}) requires that $\beta=1/2$,
such that $\lambda=1$ and
\begin{equation}
Q_{I}=-\frac{j}{k}\log 2-\log [e^F+e^{-F}]+\log [e^F+e^{-F}+2]
\label{eq:KCinfty1}
\end{equation}
which is exactly the Bhattacharyya limit \cite{viterbi}.

The average error exponent as obtained from the type II solution is given by
\begin{equation}
Q_{II}=\lambda\left[-\frac{j}{k}\log 2+\log 2\cosh[\beta F]\right]
       +\log [2\cosh(F-\beta F\lambda)]-\log 2\cosh F
\end{equation}
The condition on the entropy $\bra S_2\ket\geq 0$ is satisfied for
all $\beta>0$, whereas the condition $\bra S_1\ket\geq 0$ is violated
below the critical (freezing) temperature $1/\beta^*$ obtained from
\begin{equation}
-\frac{j}{k}\log 2-\beta^*F\tanh[\beta^*F]+\log2\cosh[\beta^*F]=0
\end{equation}
This negative entropy is an artifact of the assumption about the symmetry
between replicas, and is easily remedied by considering a `frozen RSB' ansatz
\cite{renato}. Using this ansatz and taking into account condition
(\ref{eq:condition3}), the (frozen) average error exponent obtained from
the type II solution, is finally given by
\begin{equation}
Q^{fr}_{II}=F\tanh[\beta^*F]+\frac{j}{k}\log 2-\log 2\cosh F
\label{eq:KCinfty2}
\end{equation}

What remains is to determine whether the type I or type II solution is
physically dominant, by using $Q$ as a generating function
for calculating the related free energies (through its derivative with respect
to $\lambda$).
Results for the case
of $k,j\to\infty$ are presented in figure \ref{fig:KCtoinfty} for $p=0.01$ and
$p=0.05$.

\subsection{Reliability exponent}
\label{sec:Rel_exp}

To obtain the reliability exponent we take equations
(\ref{eq:reliability_final_r})-(\ref{eq:I3_r}) and optimise with
respect to $r$.  Deriving a general set of equations similar to
(\ref{eq:SP1},\ref{eq:SP2}), that can be solved iteratively, is
difficult in this case. However, in the limit $k,j\to\infty$, we
observe that we can restrict the possible solutions of
$\hat{\pi}(\hat{x},\hat{y})$ to two different types:
\newline
Type I:
\begin{equation}
\hat{\pi}(\hat{x},\hat{y}) = \frac12
\left[\delta(\hat{x}-1)+\delta(\hat{x}+1)\right]\delta(\hat{y}-1)
\label{eq:typeIr}
\end{equation}
Type II:
\begin{equation}
\hat{\pi}(\hat{x},\hat{y}) = \frac{}{}
  \delta(\hat{x})\ \delta(\hat{y}-1)
\label{eq:typeIIr}
\end{equation}
In this case, knowledge of the solution for $\hat{\pi}(\hat{x},\hat{y})$ is
sufficient for calculation the reliability exponent
(\ref{eq:reliability_final_r}).
Furthermore, the expression obtained from the type II solution turns out to be
identical to that of the average error exponent (\ref{eq:KCinfty2}).

On the other hand, the reliability exponent obtained from the type I solution
is somewhat different, and takes the form:
\begin{equation}
Q_{r~I}=-\frac{1}{r} \frac{j}{k}\log 2-\log [\cosh(F)]+\frac{1}{r} \log
[\cosh^{r}(F) + \cosh^{r}((2\beta\lambda-1)F)] \ .
\label{eq:KCinfty1r}
\end{equation}
Given the relation (\ref{eq:condition3}) and $\beta=1/(1+\lambda)$, one obtains
$\lambda=1$, $\beta=1/2$, and the expression reduces to
\begin{equation}
Q_{r~I}=-\frac{1}{r} \frac{j}{k}\log 2-\log [\cosh(F)]+\frac{1}{r} \log
[\cosh^{r}(F) + 1] \ .
\label{eq:KCinfty1rf}
\end{equation}
Optimising the expression with respect to $r$, one obtains a similar
expression to the expurgated exponent result~\cite{gallager_book}
\begin{equation}
E_{ex}(r,R) = \mathop{\rm max}_{r} \left \{ \ln 2 \cosh F -
\frac{1}{r} \ln \left [ (2 \cosh F)^r + 1 \right ] + \frac{1}{r}(1-R)
\ln 2 \right \} \ ,
\end{equation}
%
which is also identical to the result obtained for the average bound
of the reliability exponent in~\cite{yoshiyuki}.

\begin{figure}[bh]
\setlength{\unitlength}{1.4mm}
\begin{picture}(120,75)
\put( 15,7){\epsfxsize=115mm \begin{turn}{0} \epsfbox{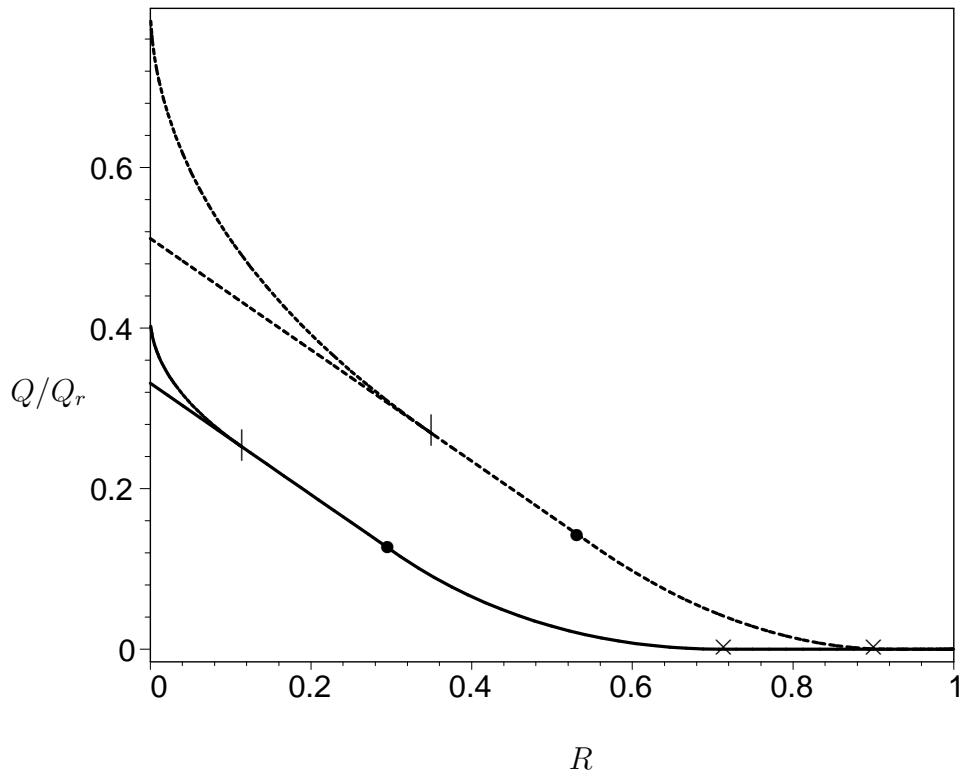} \end{turn}}
\put(7,35){$Q/Q_r$}
\put(60,0){$R$}
\put(28.5,30){$|$}
\put(46.5,31.5){$|$}
\put(60,21.5){$\bullet$}
\put(42,20.3){$\bullet$}
\put(73.5,10.8){$\times$}
\put(87.75,10.8){$\times$}
\end{picture}
\caption{Reliability and average exponents as function of the code
rate $R$ for regular $k,j\to\infty$ Gallager codes for which
analytical expressions can be derived; see (\ref{eq:KCinfty1}) and
(\ref{eq:KCinfty2}) (dashed: $p=0.01$ and solid: $p=0.05$). The
reliability exponent is identical to the average error exponent except
for very low $R$ values where it is represented by the curved
solutions above the linear average exponent results marked by a dashed
and solid lines, respectively. The transition point is marked by a
vertical line. The transition between solutions of type I and II 
is marked by $\bullet$ and the critical transition point by a $\times$.  }
\label{fig:KCtoinfty}
\end{figure}

The reliability exponent is therefore identical to the average error exponent
except for very low $R$ values as shown in figure \ref{fig:KCtoinfty} for
$p=0.01$ and $p=0.05$ (marked by a dotted line in the two cases considered).

\section{MN codes}
\label{sec:MN}

In this section, we extend our treatment of the average error and
reliability exponent to regular MN codes~\cite{MN}, a variant of
LDPC codes.

A regular MN code is defined by the binary $N\times(N+K)$ matrix
$A=[C_s|C_n]$, concatenating two sparse matrices with the $N\times
N$ matrix $C_n$ assumed invertible. The $N\times K$ matrix $C_s$
has $k$ non-zero elements per row and $j$ per column while $C_n$
has $t$ non-zero elements per row and per column. The code rate is
given by $R\equiv K/N=k/j$. The encoding scheme consists of
generating a codeword $\bt^0\in\{0,1\}^N$ from an (unbiased)
message vector $\bs^0\in\{0,1\}^K$ via $\bt^0=(C_n^{-1}C_s)\bs^0$.
Upon sending $\bt^0$ through the noisy channel the vector
$\br=\bt^0+\bn^0$ is received, where $\bn^0$ is the true channel
noise (\ref{noise}).

Decoding is carried out by multiplying the received vector $\br$
by $C_n$ to produce the syndrome vector $\bz=C_s \bs^0+C_n \bn^0$.
In order to reconstruct the original message, one selects the best
estimate $(\bs,\bn)$, for the true $(\bs^0,\bn^0)$ from the parity
check set
$\mathcal{I}_{pc}=\{(\bs,\bn)\!~|~\!C_s\bs+C_n\bn=\bz\}$, on the
basis of the message/noise statistics. Note that since we take the
message vector $\bs^0$ to be unbiased, the selection will only be
based on the noise statistics.

Since most calculation steps are completely analogous (although
lengthier) to those of Gallager codes, we only state the final
general expression for MN codes:
\begin{eqnarray}
Q_r(\beta_1,\beta_2,\lambda_1,\lambda_2)&=&\frac{1}{r}
   {\rm Extr}_{\pi,\hat{\pi},\rho,\hat{\rho}}
\left\{\frac{}{}\right.
\label{eq:reliability_final_MN_r}
\\&&\hspace*{-40mm}
   -k \log\int\left\{ d\pi(x,y)\ d\hat{\pi}(\hat{x},\hat{y})\right\}
  \left(\frac{1+x\hat{x}}{2}\right)^{r\lambda_1}
  \left(\frac{1+y\hat{y}}{2}\right)^{r\lambda_2}
\nonumber\\&&\hspace*{-40mm}
  -t \log\int\left\{ d\rho(x,y)\ d\hat{\rho}(\hat{x},\hat{y})\right\}
  \left(\frac{1+x\hat{x}}{2}\right)^{r\lambda_1}
  \left(\frac{1+y\hat{y}}{2}\right)^{r\lambda_2}
\nonumber\\&&\hspace*{-50mm}
  +\log\int\prod_{i=1}^k\left\{d\pi(x_i,y_i)\right\}
                      \prod_{l=1}^t\left\{d\rho(u_l,v_l)\right\}\!
  \left(\frac{1\!+\!\prod_i x_i\prod_l u_l}{2}\right)^{r\lambda_1}\!\!
  \left(\frac{1\!+\!\prod_i y_i\prod_l v_l}{2}\right)^{r\lambda_2}
 \nonumber\\&&\hspace*{-40mm}
  + \frac{k}{j}\log\int\prod_{c=1}^j\left\{d\hat{\pi}(\hat{x}_c,\hat{y}_c)\right\}
  \left[\sum_{\s=\pm}\prod_{c=1}^j\left(
             \frac{1+\s\hat{x}_c}{2}\right)\right]^{r\lambda_1}
  \left[\sum_{\s=\pm}\prod_{c=1}^j\left(
             \frac{1+\s\hat{y}_c}{2}\right)\right]^{r\lambda_2}
\nonumber\\&&\hspace*{-40mm}
  +\log\int\prod_{l=1}^t\left\{d\hat{\rho}(\hat{x}_l,\hat{y}_l)
  \right\}
  \left\bra\left[
  \sum_{\tau=\pm}e^{\beta_1\tau F n^0}\prod_{l=1}^t
  \left(\frac{1+\tau\hat{x}_l}{2}\right)\right]^{\lambda_1}
  \right.
\nonumber\\&&\hspace*{ 20mm}
  \left.\left.\times
  \left[
  \sum_{\tau=\pm}e^{\beta_2\tau F n^0}\prod_{l=1}^t
  \left(\frac{1+\tau\hat{y}_l}{2}\right)\right]^{\lambda_2}
  \right\ket_{n^0}^r\right\} 
\nonumber
\end{eqnarray}
with the short-hand notation $df(x,y)$=$dxdy~f(x,y)$.
As for Gallager codes, for $\beta_1=\beta_2=\beta$, and
$\lambda_1=-\lambda_2=\lambda$, $Q_r$ becomes the {\em average error exponent}
for $r=1$ , while for optimised $r$ it becomes the {\em reliability exponent}.
Furthermore, the conditions (\ref{eq:entropies}) and (\ref{eq:condition3}) must
always be satisfied.

Similarly to the case of Gallager codes one can derive a set of functional
equations (reminiscent of 'density evolution' equations \cite{Richardson}) for
$\pi,\hat{\pi},\rho$ and $\hat{\rho}$.

\subsection{Average error exponent - finite $k,j$ and $t$}
\begin{figure}[t]
\vspace{-15mm}
\setlength{\unitlength}{1.4mm}
\begin{picture}(120,55)
\put( 8,  5){\epsfysize=40\unitlength\epsfbox{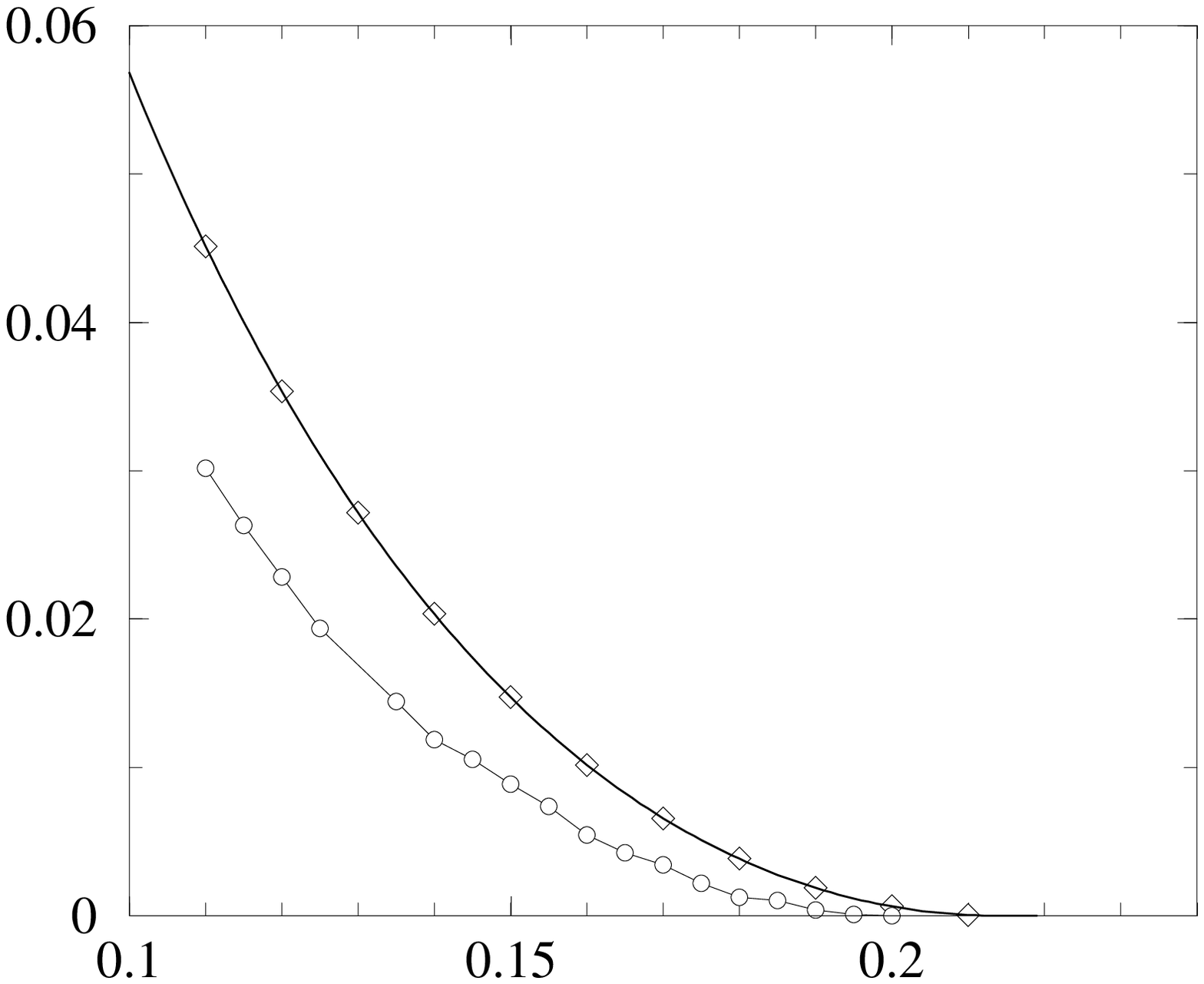}}
\put( 63,  5){\epsfysize=40\unitlength\epsfbox{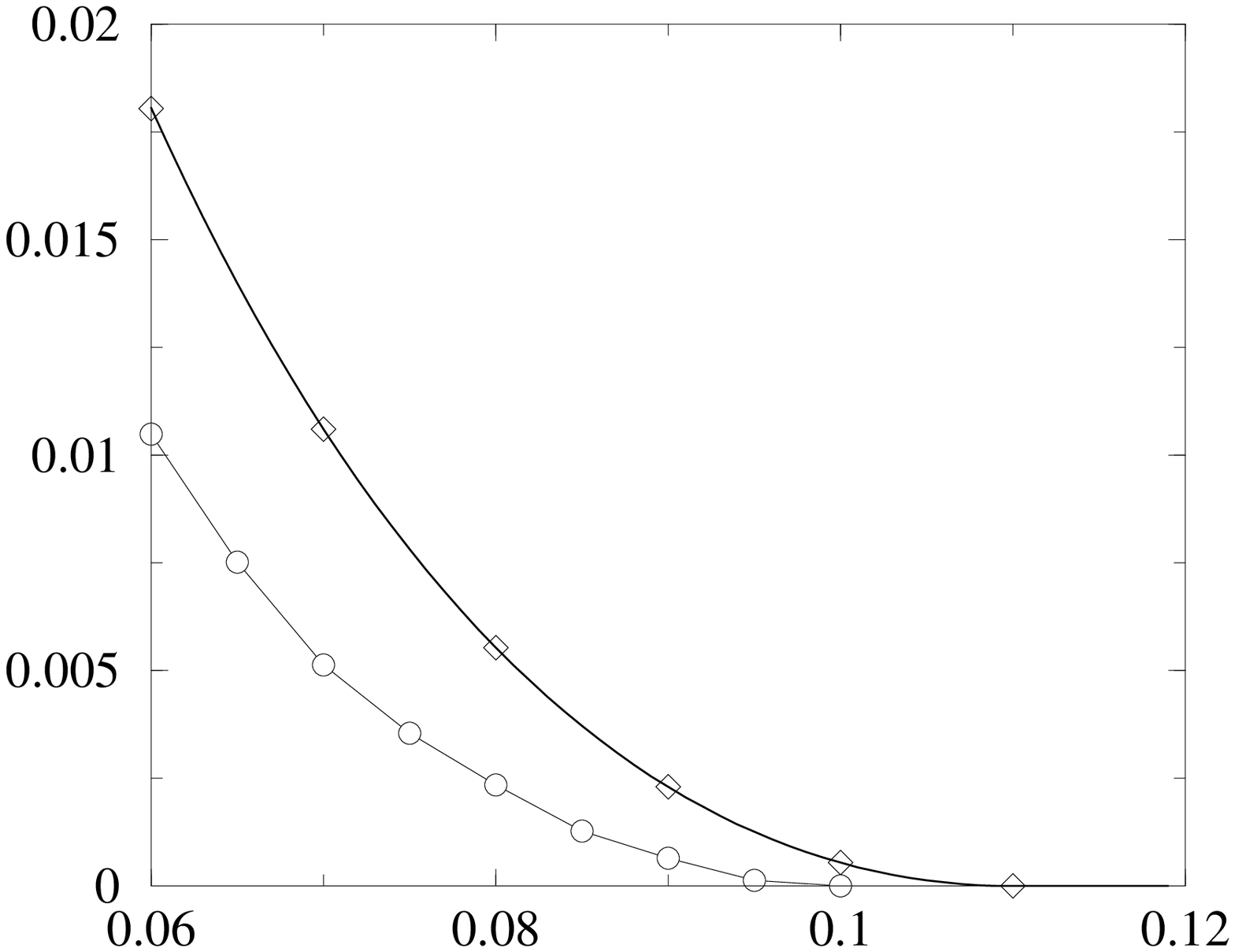}}
\put(4,25){$Q$}
\put(61,25){$Q$}
\put(35,2){$p$}
\put(90,2){$p$}
\end{picture}
\caption{Average error exponent $Q$ as function of the flip rate $p$
obtained for Gallager (\ref{eq:reliability_final_r}) and 
MN codes (\ref{eq:reliability_final_MN_r}).
Left: results for MN codes of $R=1/4$. Lower line and
circles: $(k,j,t)=(1,4,2)$.  Upper line and diamonds: $(2,8,3)$. 
Thick upper line: $k,j \rightarrow\infty$.
Right: results for Gallager and MN codes of $R=1/2$.
Lower line and circles: Gallager, $(k,j)=(6,3)$. Upper line and diamonds: MN,
$(k,j,t)=(3,6,3)$. Upper line: the analytical solution of $k,j\to\infty$ 
obtained via (\ref{eq:KCinfty1},\ref{eq:KCinfty2}).}
\label{fig:MNfinite}
\end{figure}

The average error exponent can be calculated numerically for finite
$k,j$ and $t$ values; the average error exponent $Q$ as function of
the flip rate $p$ is shown in figure~\ref{fig:MNfinite}.

On the left, we show results for MN codes of fixed rate $R=1/4$ with
three different sets of parameters $(k,j,t)=(1,4,2)$ (circles),
$(2,8,3)$ (diamonds) and $k,j \rightarrow\infty$ (upper line).
It is interesting to notice that average exponents for 
either $k > 2$ or $t > 2$ values
coincide with that of the infinite connectivity case (which can be
obtained analytically). This complements other interesting properties
of MN codes, to do with their critical flip rate values, that have
been obtained previously, distinguishing them from Gallager LDPC
codes~\cite{renato,jort,tanaka}.

On the right, we see a comparison between average error exponents of
Gallager and MN codes ($R=1/2$).  The Gallager code $(k,j)=(6,3)$
average error exponent (circles) is significantly below the random
code $k,j \rightarrow\infty$ value (thin upper line) and the
equivalent MN code $(k,j,t)=(3,6,3)$ result (diamonds).

\subsection{Average and reliability error exponents - $k,j,t\rightarrow\infty$}

The case of $k,j,t\to\infty$ is solvable exactly for all
transmission rates, and both average and reliability error
exponents can be obtained analytically. The solutions obtained as
well as the average and reliability error exponents calculated are
identical to those of Gallager LDPC codes. Retrospectively, this
is not surprising as both codes become random codes in this limit.

\section{Discussion}
\label{sec:conclusions}

In this paper we suggest a method for direct evaluation of the average
and reliability error exponent over the ensemble of LDPC
error-correcting codes of given rate and connectivity. An analytical
solution has been obtained, for both Gallager and MN codes, using
methods of statistical physics, which is in perfect agreement with
known results in the limit $k,j(,t)\to\infty$ (with $R$ finite). The
results for MN and Gallager codes become identical in this limit as
both become random codes.

Average error exponent results obtained by our method for codes of
finite $(k,j)$ values cannot be obtained using traditional approaches
used in the information theory community. As expected, they seem to be
upper bounded by the $k,j\to\infty$ curves, but suggest a profoundly
different behaviour for Gallager and MN LDPC codes.  Average error
exponent results for Gallager codes show a gradually improved
performance as the parameters ($k,j$) increase, until they finally
coincide with the $k,j\to\infty$ error exponent result. The results
for MN codes becomes identical to the $k,j\to\infty$ error exponent
result for all $k > 2$ or $t > 2$. 
To some extent, this is in agreement with
previous results obtained for the critical flip rate of MN
codes~\cite{renato,jort,tanaka} and is a result of the close-to-random
codebook they generate.

An interesting feature of the present study is the similarity of our
equations to those obtained in~\cite{yoshiyuki} in spite of the
different approaches used. An important advantage offered by the
current approach is a potential extension to select high performance
codes to obtain {\em reliability exponent} values for LDPC codes of
finite connectivity; obtaining such solutions remains a difficult task
and is currently under study.

\ack Support from EPSRC research grant GR/N63178 (DS,NS), the Royal
  Society (DS,NS,JvM), Grants-in-aid, MEXT (13680400 and 13780208) and
  JSPS (YK) is acknowledged.

\section*{Reference}


\begin{thebibliography}{99}
\bibitem{Richardson} T Richardson, A Shokrollahi and R Urbanke (2001),
\emph{IEEE Trans. on Info. Theory} \textbf{47} 619-637
\bibitem{renato}
R Vicente, D Saad and Y Kabashima (1999)
\emph{Phys Rev E} \textbf{60} 5352-5366
\bibitem{irregular}
R Vicente, D Saad and Y Kabashima (2000)
\emph{J Phys A} \textbf{33} 6527-6542
\bibitem{jort}
J van Mourik, D Saad and Y Kabashima (2002)
\emph{Phys Rev E} \textbf{66} 026705
\bibitem{tanaka}
T Tanaka and D Saad, (2003)
submitted in \emph{Phys Rev E}
\bibitem{yoshiyuki}
Y Kabashima, N Sazuka, K Nakamura and D Saad (2001)
\emph{Phys Rev E} (2001) \textbf{64} 046113
\bibitem{mont_sourlas}
A Montanari and N Sourlas
\emph{Eur Phys J B} \textbf {18} 107-119
\bibitem{montanari}
A Montanari
\emph{Eur Phys J B} \textbf{23} 121-136
\bibitem{franz} S.~Franz, M.~Leone, A.~Montanari and
F.~Ricci-Tersenghi, \emph{Phys Rev E} \textbf{66} 046120
\bibitem{gallager_book} R G Gallager (2001) 'Information theory and
  reliable communication' Wiley \& Sons, NY
\bibitem{viterbi}
A J Viterbi and J K Omura (1979) `Principles of Digital
Communication and Coding', McGraw-Hill Int Ed (Singapore)
\bibitem{MN} D.J.C.~MacKay (1999) \emph{IEEE Trans.\@ Inf.\@ Th.\@} \textbf{45}
399
\bibitem{miller} G.~Miller and D.~Burshtein (2001)
\emph{IEEE Trans.~Infor.~Theory} \textbf{47} 2696
\bibitem{nishimori_book} H Nishimori (2001)
'Statistical Physics of Spin Glasses and Information Processing',
Oxford University Press, UK
\bibitem{wong-sherrington} K Y M Wong
and D Sherrington (1987)
\emph{J Phys A} \textbf{20} L793-L799

\end{thebibliography}
\end{document}